\documentclass[12pt,preprint]{aastex}

\usepackage{lscape}

\def\farcs{\hbox{$.\!\!^{\prime\prime}$}}

\def\half{{\leavevmode\kern.1em\raise.5ex
\hbox{\the\scriptfont0 1}\kern-.1em /
\kern-.15em\lower.25ex\hbox{\the\scriptfont02}}} 
\def\gtsim{\lower.5ex\hbox{$\buildrel > \over\sim$}}
\def\ltsim{\lower.5ex\hbox{$\buildrel < \over\sim$}}

\slugcomment{To appear in The Astronomical Journal}

\shorttitle{X-Rays from IRAS$\,$20126}
\shortauthors{Anderson et al.}

\begin{document}

\title{X-Ray Emission from Young Stars in the Massive Star Forming Region IRAS$\,$20126+4104}

\author{C. N. Anderson$^1$, P. Hofner$^{1,2}$, D. Shepherd$^2$ and M. Creech-Eakman$^{1,2}$}

\affil{$^1$Physics Department, New Mexico Tech, 801 Leroy Pl., Socorro, NM 87801, USA}

\affil{$^2$National Radio Astronomy Observatory, P.O. Box O, Socorro, NM 87801, USA}

\begin{abstract}

We present a $40\,$ks Chandra observation of the IRAS$\,$20126+4104 core region. In the inner
$6^{\prime\prime}$ two X-ray sources were detected, which are coincident with the radio jet source I20S
and the variable radio source I20Var. No X-ray emission was detected from the nearby massive protostar I20N.
The spectra of both detected sources are hard and highly absorbed, with no emission below $3\,$keV.  

For I20S, the measured $0.5-8\,$keV count rate was $4.3\,$cts$\,$ks$^{-1}$. 
The X-ray spectrum was fit with an absorbed 1T APEC model with an energy
of kT$\,=10\,$keV and an absorbing column of 
N$_H = 1.2\times 10^{23}\,$cm$^{-2}$. An
unabsorbed X-ray luminosity of about $1.4\times 10^{32}\,$erg$\,$s$^{-1}$ was estimated.
The spectrum shows broad line emission
between 6.4 and 6.7\, keV, indicative of emission from both neutral and highly ionized iron. The X-ray
lightcurve indicates that I20S is marginally variable; however, no flare emission was observed.

The variable radio source I20Var was detected with a count rate of $0.9\,$cts$\,$ks$^{-1}$ but there was no evidence
of X-ray variability. The best fit spectral model is a 1T APEC model  
with an absorbing hydrogen column of N$_H = 1.1\times 10^{23}\,$cm$^{-2}$ and
a plasma energy of kT = 6.0$\,$keV. The unabsorbed X-ray luminosity is about
$3\times 10^{31}\,$erg$\,$s$^{-1}$.
\\
\\

\end{abstract}

\keywords{Individual: IRAS 20126+4104, stars: emission-line, stars: formation, stars: pre-main sequence, X-rays: general, X-rays: stars}

\newpage

\section{Introduction}

Massive protostars are still embedded in their natal molecular material, and due to
the high extinction are traditionally observed in the radio/mm and infrared wavelength bands.
X-rays with energy $> 2\,$keV can also penetrate dense molecular cloud cores, and recent observations 
have shown that X-ray emission can be detected from regions of massive star formation
(e.g. Hofner et al. 1997, 2002, Feigelson et al. 2005, Getman et al. 2006). In particular, the arc-second resolution imaging capability of 
the Chandra X-ray Observatory has allowed for the detection of young, and hence embedded, massive stars
in a number of the nearest massive star forming regions (e.g. Townsley 2006).  
Here we present the results of Chandra observations of the IRAS$\,$20126+4104 region, which contains
a massive protostar, i.e. a massive star that is still in the process of assembling
most of its final mass through accretion from its envelope. A small number of low-mass `class 0' type protostars have
been detected in X-rays (e.g. Tsuboi et al. 2001) and the observations presented in this paper were carried out
to search for X-ray emission from similar, but more massive objects.
  
The IRAS$\,$20126+4104 region of massive star formation has been studied extensively in the last few 
years. It is relatively nearby (1.7$\,$kpc), has a FIR luminosity of 
about $10^4\,$L$_\odot$, and shows classic signs of massive star formation: dense and hot molecular gas
(e.g. Cesaroni et al. 1997);  maser emission from the H$_2$O, OH, CH$_3$OH, and NH$_3$ 
molecules (e.g. Trinidad et al. 2005, Edris et al. 2005, Kurtz et al. 2004, Zhang et al. 1999); and a massive molecular flow (e.g. Su et al. 2007, Shepherd et al. 2000). The radio continuum emission
in IRAS$\,$20126+4104 is very weak (Hofner et al. 1997), which demonstrates the early evolutionary 
state of the region. In the central core, a disk/jet system was observed and the dynamics of the disk
indicated a protostar  of mass $\approx 7\,$M$_\odot$, which is thought to be accreting mass through
the disk (e.g. Cesaroni et al. 1999, 2005). Infrared observations support this picture, but also show
that the inner $0.1\,$pc of the IRAS$\,$20126+4104 region hosts a complex environment with multiple stars
and different flow components (e.g. Shridaran et al. 2005, De Buizer 2007).

Hofner et al. (2007) studied the radio continuum emission in the IRAS$\,$20126+4104 core and found
3 compact sources: I20N1 and I20N2,  and I20S, located about $1^{\prime\prime}$ to the south.
Figure~1 shows the $3.6\,$cm continuum map of this region. 
These authors discuss the possible origin of the radio continuum sources, and while the detailed nature of the radio continuum  sources
is still under debate, they favor a model for I20N1 and N2 where the ionization 
is shock-induced by the outflow from the putative massive proto-star, located about $0\farcs3$ to the south-east of I20N1.
For I20S, on the other hand, the radio continuum data are consistent with direct photo-ionization of jet material from an
early B-type star.

Several models have been proposed to explain X-ray emission from early type stars: 
shocks in line driven winds in the outer atmosphere of the star (e.g., Lucy \& White 1980), magnetically confined stellar winds
(Gagn\'e et al. 2005), collisions of strong winds in close binary systems (e.g., Pittard \& Parkin 2010), and binary induced magnetic reconnection events
(Schultz et al. 2008). For massive stars in very early evolutionary states we also need to consider X-ray emission from accretion events
(e.g., G\"unther et al. 2007), as well as interaction of outflows/jets with the surrounding medium (e.g., G\"udel et al. 2005, Pravdo, Tsuboi \& Maeda 2004). 
In this paper we present
Chandra observations of IRAS$\,$20126+4104 with the goal to search for and investigate these X-ray emission processes from massive
protostars.

In section 2 of this paper, we describe the Chandra observations and data reduction methods.
Section 3 presents the observational results, which are further discussed in section~4. We summarize
the paper in section 5.

\section{Observations and Data Reduction}

The IRAS$\,$20126+4104 region of massive star formation was observed
with the Advanced CCD Imaging Spectrometer (ACIS) on board the
Chandra X-Ray Observatory on March 17, 2003.
The energy range of ACIS is 0.1 to 10\,keV and the total exposure time was
$39.35\,$ks.     For details on the instrument see 
Weisskopf et al. (1996), Weisskopf et al. (2002), and
Garmire et al. (2003).   The nominal pointing position for the
ACIS array was R.A. (J2000) = $20^h14^m30\fs27$, Dec.
(J2000) = $+41^\circ13\arcmin42\farcs1$.   The observations were taken in
the standard ``Timed Event, Very Faint'' telemetry mode.
The roll angle of the space craft during the observations was $58.19^\circ$, and the focal plane
temperature was $-119.6\,^\circ$C.   
Although 6 CCD chips (I0-I3, S2, S3)
were active during the observations, no useful data were obtained from the
spectroscopic array and we report here only data from the imaging array, 
ACIS-I.   The imaging array consists of four abutted $1024\times 1024$ pixel CCDs
(pixelsize $0\farcs492$) covering an angular region of about $17^\prime \times 17^\prime$.
Data reduction was performed using the CIAO software package version 
3.3.01 provided by the Chandra X-ray Center, starting from level~2
reprocessed data (processing version DS 7.6.8). This version of the data
processing pipeline provides an improved aspect solution and
correction of effects due to the increase of the charge transfer inefficiency
(Townsley et al. 2000). ASCA grades 0, 2, 3, 4, 6, were selected and 
the data were gain-corrected and filtered for bad CCD pixels and times of bad
aspect. The energy range was restricted to $0.5 - 8\,$keV, where the
point spread function was of good quality.  
Exposure maps were created and applied to the data in the standard fashion.
No background flares were detected during the observations and the average
background emission, as measured in a source-free region in the ACIS-I chips,
was $2.3\times 10^{-7}\,$count$\,$s$^{-1}\,$pixel$^{-1}$.

To check the astrometric accuracy of the Chandra data, the positions of $17$ bright
X-ray sources located near the center of the ACIS array were compared with their counterparts in the Two 
Micron All Sky Survey (2MASS, Skrutskie et al. 2006)\footnote{http://www.ipac.caltech.edu/2mass/} 
catalogue. Although the initial astrometry was already very good, with a maximum individual deviation of 
$< 0\farcs4$ in either coordinate, a small systematic offset of $0\farcs2$ toward the west and 
$0\farcs1$ toward the south was detected. After correction of the Chandra source positions, the root-mean-square (RMS) offset in R.A. and Dec. between Chandra and 2MASS was about $0\farcs1$. We take this 
value as an indication of the astrometric accuracy of our data.

\section{Results and Analysis}
\subsection{Source Detection}

Sources were identified in the ACIS-I field-of-view using WAVDETECT, a wavelet-based source detection 
program that works well to detect closely spaced sources (Freeman et al. 2002).  We used a  ``threshold 
significance'' of $10^{-6}$ and wavelet scale sizes from 1 to 16 pixels incremented by a factor $\sqrt{2}$. 
These values provided good sensitivity to faint sources (e.g. $< 100$\,counts).
To more reliably identify weak sources with emission only in the soft ($0.5-2\,$keV) or hard ($2-8\,$keV) X-ray 
energy ranges, WAVDETECT was run for each range separately as well as for the full energy range 
($0.5-8\,$keV). All sources found by WAVDETECT were inspected visually. We found that spurious
detections occured below 7 counts, often having the appearance of linear stripes and edge effects.
These sources were removed from our list of detected sources. Sources 
with at least 7 counts within the source detection region identified by WAVDETECT were considered to be 
real detections.

The total number of sources detected was 150; this included sources that were only detected in soft or hard 
X-rays as well as those that were detected in the full energy range.  The brightest detected X-ray source 
had a total of 340 
counts and was identified with a foreground Main Sequence K05 spectral type star.  The second brightest 
source (167 counts) corresponds
to the radio source I20S.
Approximately $80\,\%$ of the sources had less than $50$ counts.
In this paper we focus on the two  X-ray sources detected in the central $6''$ of the IRAS$\,
20126+4104$ region.
The entire data set of the X-ray cluster in IRAS$\,20126+4104$ will be presented in a subsequent paper.

A gray scale image of the X-ray emission detected in the central  $6^{\prime\prime}$ is shown in Figure~2.
The peak positions of the $3.6\,$cm radio continuum sources from Hofner et al. (2007) are shown as crosses.
Two sources are clearly detected by Chandra.  The bright X-ray source 
CXO$\,$J201426.0+411331.7 with 167 counts within $2-8\,$keV, or a count rate of
$4.3\,$cts$\,$ks$^{-1}$, is coincident with the peak 
 position of the ionized radio jet source I20S.
 The X-ray emission from I20S is consistent with a point-source, and its J2000 
 coordinates are R.A.(J2000) = $20^h14^m26\fs03$, Dec.
(J2000) = $+41^\circ13\arcmin31\farcs7$. 
 Due to the highly accurate astrometry ($\approx 0\farcs1$) of both radio and X-ray data, 
 there can be little doubt that the X-ray emission is associated with the radio jet source I20S. 
There is no indication of X-ray emission from the
massive protostellar candidate I20N, which
is located about $1^{\prime\prime}$ to the north. 
 

 The second X-ray source detected in the inner $6''$, CXO$\,$J201426.2+411327.9, has a total of 33 counts 
 (count rate of $0.9\,$cts$\,$ks$^{-1}$) and, as with I20S, is 
 only detected in the hard energy range. The emission is point-like, and its position is:
 R.A.(J2000) = $20^h14^m26\fs25$, Dec.
(J2000) = $+41^\circ13\arcmin27\farcs9$.  
 This source is coincident with the variable radio source I20Var (Hofner et al. 2007). 
 
\subsection{Timing Analysis}

Timing analysis was performed to determine whether either source displayed X-ray variability using the XRONOS software 
package\footnote{http://heasarc.gsfc.nasa.gov/docs/software/lheasoft/xanadu/xronos}. 
For I20S and I20Var, the count rate versus time (i.e. X-ray lightcurve) was determined by
measuring counts in 2000\,s temporal bins
within angular regions defined by WAVDETECT.  The background from a nearby, source-free region was 
then subtracted to obtain the final lightcurve. Analysis of the lightcurves was done using the LCSTATS 
program in the XRONOS package. 
Due to the low number of counts in our observation, the $\chi^{2}$ 
method was used to determine source variability.
Lightcurves for I20S and I20Var are shown in Figure~3.

The X-ray emission from I20S (Figure~3, upper panel) was found to be marginally variable, with a  $\chi^{2}$ 
probability for constancy of $1.1\times 10^{-5}$. The variation of the X-ray flux in this source appears to be 
smooth on a time scale of a few times $10^4\,$s, and
it did not exhibit any strong flare-like behavior.

For I20Var (Figure~3, lower panel) the  $\chi^{2}$ probability for constancy was 0.4, indicating no variation of its X-ray 
flux was detected.  
I20Var has shown strong variability in 3.6\,cm radio continuum observation,
where it exhibited a flux density increase of a factor of 40 within 2 observations taken between 1998 and 
2000 (Hofner et al. 2007). 
Since the 3.6\,cm and X-ray observations were not simultaneous, it is possible 
that the X-ray emission was simply quiescent at the time of the observations.
In the I20Var lightcurve, the last bin shows a sudden rise to more than 3 times the average 
count rate, which might indicate the onset of an X-ray flare. Nearby X-ray sources on the ACIS-I chip do not
show any similar behavior, hence it is unlikely that this is an instrumental effect.

\subsection{Spectroscopy}

The $0.5 - 8\,$keV spectra of I20S and I20Var are shown in Figure 4. Both sources are
highly absorbed with no X-ray counts detected below $3\,$keV. 
Model fitting of the I20S and I20Var spectra was performed using the XSPEC\footnote
{http://heasarc.gsfc.nasa.gov/docs/software/lheasoft/xanadu/xspec} software package. 

In order to fit the X-ray spectrum of I20S, we tried a variety of models including optically thin, 
one (1T) and two-temperature (2T) thermal models (APEC\footnote{APEC code v1.3.1, http://hea-
www.harvard.edu/APEC}) and mixed thermal and non-thermal models with fixed solar 
abundance of 0.2. All models were also run allowing the abundance to vary.
None of these models fit the data well, however all models imply similar absorption columns 
of about $10^{23}\,$cm$^{-2}$, as well as the presence of a hot plasma with kT$\,\ge\, 6\,$keV.

In Figure~4 (upper panel), we show an overlay of our best fit model, which consists of an absorbed 1T APEC model plus three gaussians.
The best fit model parameters are an absorbing column of  N$_H = 1.2^{+0.3}_{-0.2}\times 10^{23}\,$cm$^{-2}$,
a plasma energy of kT$\,=10^{+9}_{-2}\,$keV, and an abundance of 1.2 solar.
Two of the three gaussians lines were added to represent energies of 3.9 and $4.9\,$keV,
which are likely due to Ca$\,$XIX. Whether in fact these lines are actually present in the
data cannot be stated with any certainty due to the low signal-to-noise ratio, but their
presence improves the fit and similar
emission lines have been reported in the class I protostar WL22 in $\rho$-Ophiuchi
(Imanishi \& Koyama 2001; see also Maggio et al. 2007).
At the plasma temperature kT$\,=10^{+9}_{-2}\,$keV, emission from the Fe$\,$XXV (He-like) ion
is expected at an energy of $6.7\,$keV, and 
the spectrum of I20S shows excess emission near this energy. However, the emission
appears quite broad, and we add the third gaussian at an energy of $6.4\,$keV, which
represents emission from weakly ionized iron. 

While I20S is the second brightest X-ray source in the 
IRAS$\,$2016+4104 region, its emission is still fairly weak and provides only a total of 167
X-ray photons, which does not allow strong constraints on spectral modeling.
We thus do not consider the model presented above as fully representative of the physical
state of the X-ray emitting plasma in I20S. The most significant results of the
model fit are the absorption column density, as well as the broad emission complex around
the iron line energies, implying emission from both weakly ionized, and He-like iron. The observed
(i.e. absorbed)  $0.5 - 8\,$keV flux of I20S is $1.3\times 10^{-13}\,$erg$\,$cm$^{-2}\,$s$^{-1}$,
and the unabsorbed luminosity is about  $1.4\times 10^{32}\,$erg$\,$s$^{-1}$.

The X-ray spectrum of I20Var is shown in Figure~4 (lower panel). The low count rate precludes a
detailed spectral analysis. A 1T APEC model with a fixed abundance (0.2 solar) was fit to the X-ray data. 
Our best fit has an absorbing hydrogen column of N$_H = 1.1\times 10^{23}\,$cm$^{-2}$, and
a plasma energy of kT$=6.0\,$keV. We estimate that N$_H$ is accurate within about a factor of 2,
but the plasma temperature is less well constrained; reasonable fits were possible
with any energy larger than about $2.3\,$keV. The observed 
$0.5 - 8\,$keV flux of I20Var is $2.7\times 10^{-14}\,$erg$\,$cm$^{-2}\,$s$^{-1}$, corresponding
to an unabsorbed luminosity of approximately $3\times 10^{31}\,$erg$\,$s$^{-1}$.

\section{Discussion}

\subsection{I20N}
 
We first turn to the non-detection of X-rays from the position of the massive protostar I20N.
As has been pointed out by Skinner et al. (2007), very few massive embedded stars have been detected in 
X-rays and little is known about their X-ray emission during early evolutionary stages.
Among the candidates for massive protostars which have been detected are the well
known sources HH80-81 (GGD27-X, Pravdo et al. 2009), S106-IRS4 (Giardino et al. 2004),
NGC2071-IRS1, IRS3 (Skinner et al. 2009),
and several sources in the MonR2 region (Kohno, Koyama \& Hamaguchi et al. 2002).
These sources show relatively hard spectra with energies in the range of $2 - 8\,$keV, observed
through column densities $10^{22} - 10^{23}\,$cm$^{-2}$ and unabsorbed X-ray luminosities of
$10^{30} - 10^{31}\,$erg$\,$s$^{-1}$. Translating our detection limit of 7 counts to physical parameters
is model dependent: for an optically thin thermal spectrum with an energy of $3.5\,$keV, and 
absorption column of N$_H = 10^{22}\,$cm$^{-2}$, our detection limit corresponds to an
X-ray luminosity of about $1\times 10^{30}\,$erg$\,$s$^{-1}$ in the $0.5 - 8\,$keV energy band.
As the column density increases, much more luminous sources would be required for detection, e.g.
L$_X = 7\times 10^{30}\,$erg$\,$s$^{-1}$ for N$_H = 10^{23}\,$cm$^{-2}$, and
L$_X = 2\times 10^{32}\,$erg$\,$s$^{-1}$ for N$_H = 10^{24}\,$cm$^{-2}$.
I20N is located near the column density maximum in the IRAS$\,$20126+4104 core region, where
Cesaroni et al. (1999) estimate N$_H \approx 10^{25}\,$cm$^{-2}$. Thus, at present, 
the non-detection of I20N in X-rays is consistent with the X-ray properties of known massive protostellar
candidates if the hypothetical X-ray
source in I20N suffers the full extent of the absorption through the column density of 10$^{25}$ cm$^{-2}$.

\subsection{I20S}
  
This source is located about $1^{\prime\prime}$ south of I20N near the edge of the dense central core.
No compact emission in any molecular transition or dust continuum has been reported at this
position; the only detection so far is at $3.6\,$cm (Hofner et al. 2007). At this wavelength, the
source shows a smooth elongated structure of axis ratio $\approx 5$ oriented in the direction of the
large scale flow (Figure~1), which has been interpreted by Hofner et al. (2007) as either a photoionized
jet or shock ionization due to outflowing matter.
 

The detailed evolutionary state of I20S is difficult to assess. Most existing models for jets are based on
collimated disk winds (e.g. Pudritz et al. 2007), hence the presence of a radio jet inside
the dense molecular cloud core suggests that an accretion disk might exist around the central object in I20S. 
However, the absence of any
detectable, compact, high density tracer suggests that the total mass of such a putative disk, and the accretion rate would be substantially lower than
in I20N. If the radio continuum emission from I20S is due to ionizing photons beamed into the
solid angle of the jet, the required UV photon flux corresponds to a ZAMS spectral type B1
(Hofner et al. 2007). This spectral type is likely an upper limit since a high velocity jet
might produce a significant fraction of the ionizing radiation. Accretion onto the stellar surface of such an object even at free-fall
velocities would result in a spectrum with E$\, < 2\,$keV (G\"unther et al. 2007), i.e. much softer than is observed.

One scenario which can produce plasmas with energies
of several keV is strong shocks, which can occur when the outflowing matter in
the jet is stopped by the surrounding dense gas in the molecular core
(e.g. Raga, Noriega-Crespo \& Velazquez 2002). As discussed above, the
energy of the X-ray emitting plasma of I20S is poorly determined, but it is likely larger than $5\,$keV.
Shock speeds $ > 2000\,$km$\,$s$^{-1}$ are required to produce a plasma of this energy.
Terminal wind speeds of early B stars can be of this magnitude (e.g. Cassinelli et al. 
1994), and proper motions within the HH80-81 jet
as large as $1400\,$km$\,$s$^{-1}$ have been detected (Marti, Rodriguez \& Reipurth 1995), so 
this scenario is a distinct possibility. The radio morphology of I20 is reminiscent of the jet in the classical T-Tauri star DG Tau which is also known to be an
X-ray emitter (G\"udel etal. 2005, 2008).
However the X-ray emission from I20S arises from a much hotter plasma, suffering much higher extinction, and is more luminous than DG Tau, so that
I20S might be an upscaled version of DG Tau.

For lower mass stars, hard X-rays are produced mostly by magnetic reconnection events which heat the
plasma to high temperatures (e.g. Feigelson \& Montmerle 1999), and similar processes have recently
been detected for massive stars (Schulz et al. 2006). This type of emission usually occurs in X-ray
outbursts with a typical flare profile in the lightcurve.
While there is evidence for marginal variability of the X-ray flux from I20S, no flare-type
variability is seen. This type of flux variation is more likely
related either to the rotation of the star or to orbital modulation in a binary. For the former
case, a possible production mechanism for hard X-ray emission would be the magnetically confined
wind model (Gagne et al. 2005), and for the latter case, the colliding binary
wind model (e.g. Pittard \& Parkin 2010). 
 
We now turn to a discussion of the emission line at $6.4\,$keV.  This line is due to weakly
ionized iron indicating the presence of relatively cold gas. Following the theory outlined in Tsujimoto et al.
(2005), if the fluorescent matter is along the line of sight to I20S, an order of magnitude estimate
for the equivalent width (EW) of the $6.4\,$keV line is
EW$\, \approx\,$(N$\,^{\prime}_H$/$10^{22}\,$cm$^{-2})\,$eV,
where N$\,^{\prime}_H$ is the equivalent hydrogen column in the fluorescent material.
The absorption column measured from model fitting the I20S X-ray spectrum is
N$_H =1.1 \times 10^{23}\,$cm$^{-2}$.  
Based on the IRAM Plateau de Bure $1.3\,$mm observations  ($\theta_{syn} = 0\farcs6$)
of Cesaroni et al. (1999), and the formulas of Mezger et al. (1990) with an assumed dust 
temperature of $100\,$K, we estimate a column density of N$_H =3 \times 10^{23}\,$cm$^{-2}$
at the position of I20S. Considering the uncertainties (angular resolution, dust properties and assumed 
temperature), this value is consistent with the absorption derived from the X-ray spectrum.
Thus, the expected EW of the $6.4\,$keV iron line is on the order of a few times $10\,$eV, whereas
much larger values for EW are necessary to explain the excess in the spectrum near $6.4\,$keV, suggesting
a much higher column density for the fluorescent matter.
As noted above, I20S is located near the edge of
the dense molecular core, which contains the massive protostar I20N. A likely scenario is therefore that the
$6.4\,$keV line arises from reflection of X-ray emission from I20S at the edge of the dense molecular core.  
However the low count-rate of the I20S spectrum precludes a more detailed analysis and more sensitive
observations are needed to study the putative $6.4\,$keV line.

\subsection{I20Var}

This source was first detected by Hofner et al. (2007) at radio wavelengths, where
it showed strong variability on time scales of $20$ days or longer.  The emission characteristics of I20Var are
consistent with gyro-synchrotron emission from a low mass, pre-main sequence star, which infrared
emission is made undetectable by an overlaying column of about N$_H = 10^{23}\,$cm$^{-2}$.
Our X-ray data imply the same amount of absorption from fitting the low energy cutoff.

With the exception of a possible flare onset at the
end of these observations, the emission from I20Var appears mostly quiescent (Figure~3, bottom) rather than
flare-like. The luminosity of quiescent X-ray emission from low mass pre-main sequence stars
is generally found to be smaller than a few times $10^{30}\,$erg$\,$s$^{-1}$ (e.g. Imanishi et al. 2001),
whereas I20Var has an unabsorbed X-ray luminosity of approximately $3\times 10^{31}\,$erg$\,$s$^{-1}$.
This suggests that I20Var is either a low mass star with unusually high quiescent X-ray emission,
or that it is an intermediate mass star. 

\section{Summary}

In this paper, we described Chandra observations of the IRAS$\,$20126+4104 core region.
The massive protostar I20N was not detected in our observations. If the X-ray luminosity of this
source is similar to the massive protostars detected so far, then the non-detection indicates
that the X-ray emission suffers the full extent of the absorption toward I20N, which is thought
to be dominated by an edge-on accretion disk. X-ray emission was detected from the radio jet source I20S. The source was marginally variable, but showed no evidence of X-ray flares. The emission has a
hard and strongly absorbed spectrum, showing a broad line between 6.4 and 6.7\,keV. The line
is consistent with a superposition of emission from both weakly, and highly ionized iron.
Also, X-ray emission from the variable radio source I20Var was detected. The X-ray emission from
this source was constant during the observations, with a larger X-ray luminosity than 
would be expected from a low mass pre-main sequence star. Together with detection
limits from 2MASS, this suggests that I20Var is possibly an intermediate mass star. 

\acknowledgements

 We would like to thank E. Feigelson, L. Townsley and J. Eilek for helpful discussions.
 This project was supported by Chandra Award No. GO3-4018A and the New Mexico Space Grant 
 Consortium.  PH and MCE also acknowledge partial support from NSF grant AST-0908901. PH and MCE are also Adjunct Astronomers at the National Radio Astronomy Observatory. We also thank the anonymous referee for comments which improved the manuscript.

\clearpage

\includegraphics{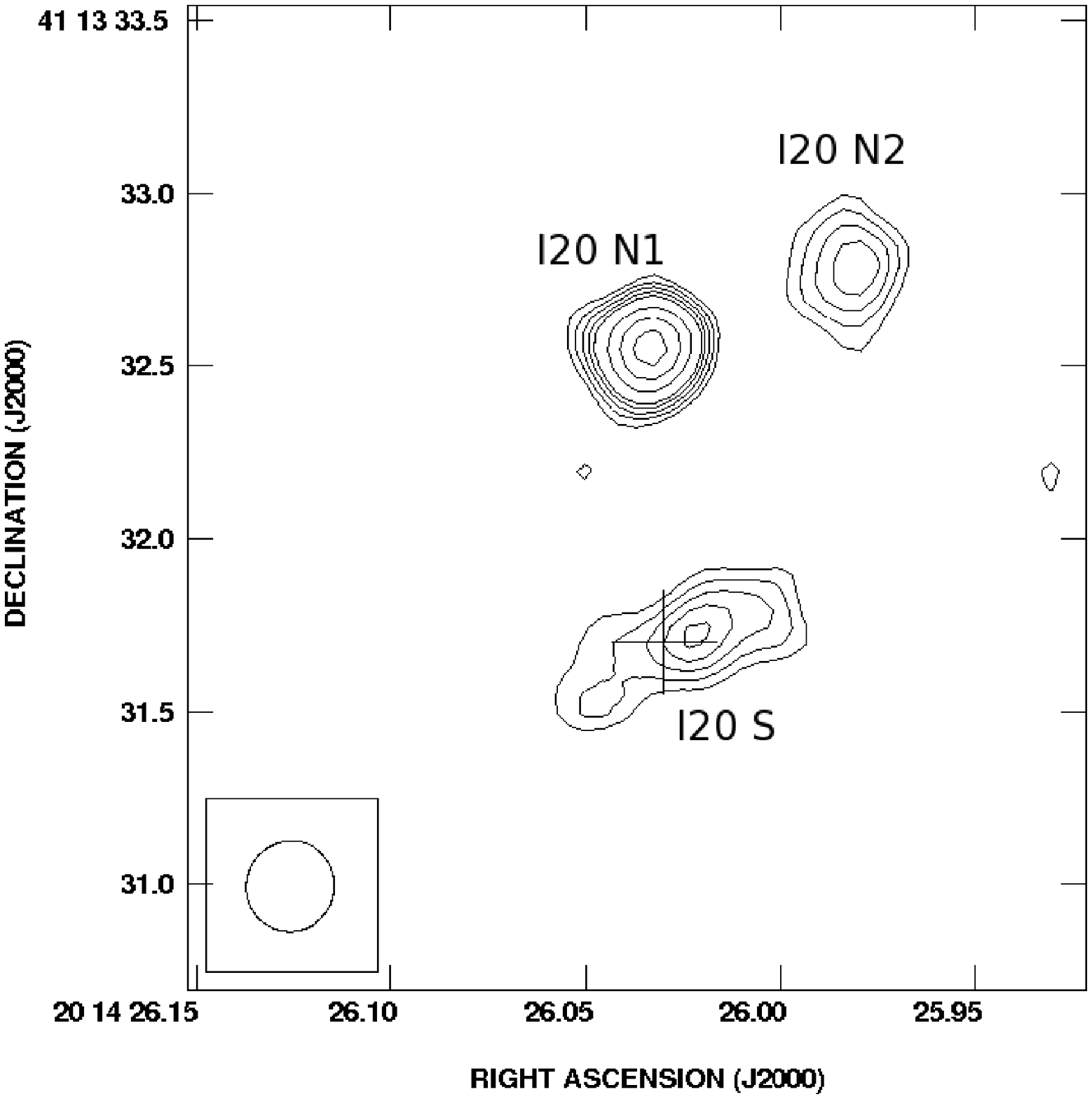}

\hspace*{0.0cm}
\begin{minipage}{15cm}
\vspace{14cm}
{\small Figure~1. Radio continuum emission at $3.6\,$cm toward IRAS$\,$20126+4104 is
shown in contours (Hofner et al. 2007).
Contour levels are -3, 3, 4, 5, 6, 7, 10, 13,16 $\times  \,8\,\mu$Jy/beam. The size of the synthesized
beam is shown in the lower left corner. The cross
marks the position of the X-ray source CXO$\,$J201426.0+411331.7 associated with I20S.
The size of the cross is 3 times the positional accuracy of our
X-ray data.}
\end{minipage}

\clearpage

\includegraphics{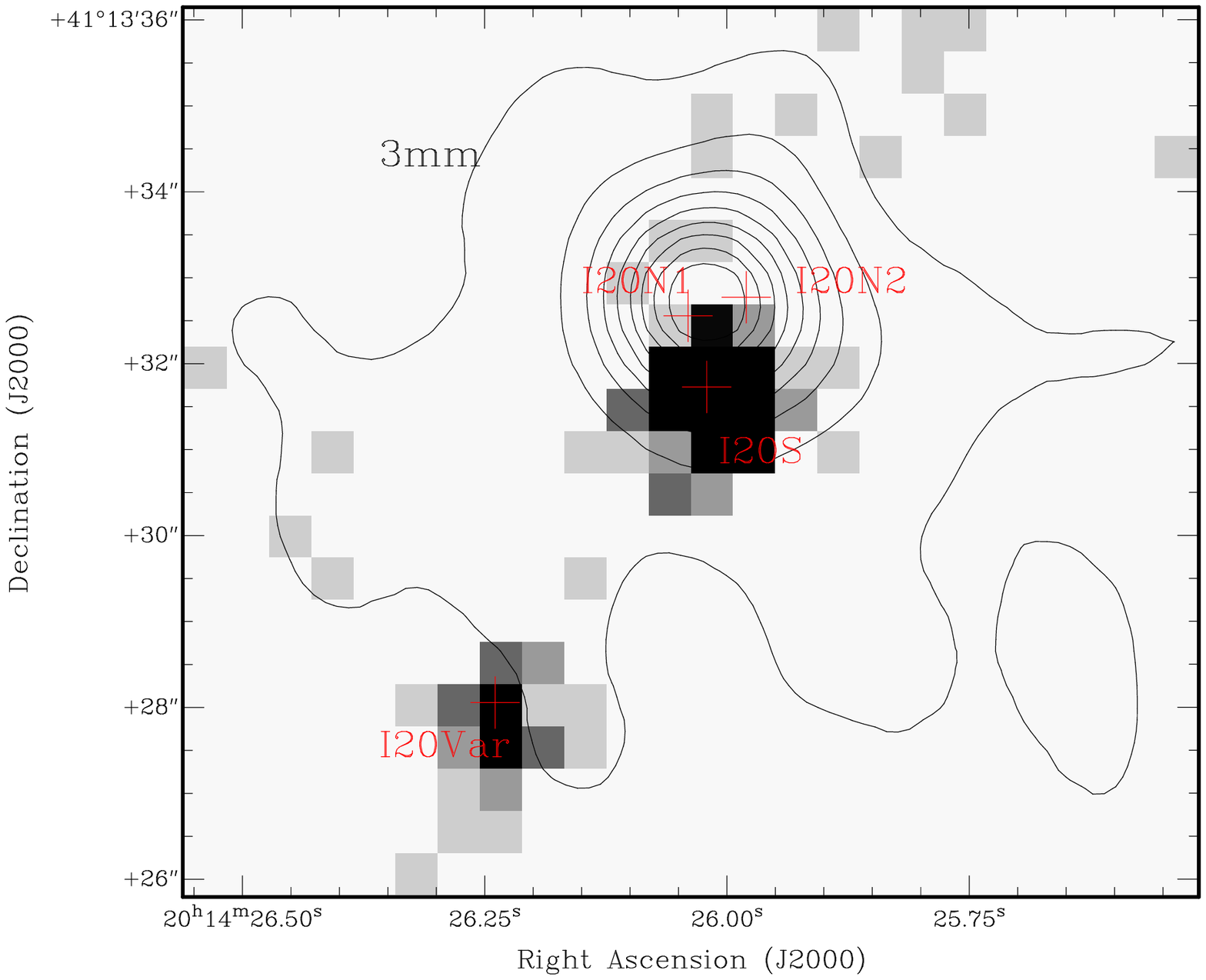}

\hspace*{0.0cm}
\begin{minipage}{15cm}
\vspace{12.5cm}
{\small Figure~2. The $0.5 - 8\,$keV X-ray emission in the IRAS$\,$20126 core region
is shown in grey scale overlayed on the $3\,$mm continuum emission
from Cesaroni et al. (1999). The red crosses show the peak positions of
the $3.6\,$cm continuum sources from Hofner et al. (2007), with the size of the crosses six times the 
astrometric error of the VLA observations.}
\end{minipage}

\clearpage

\includegraphics{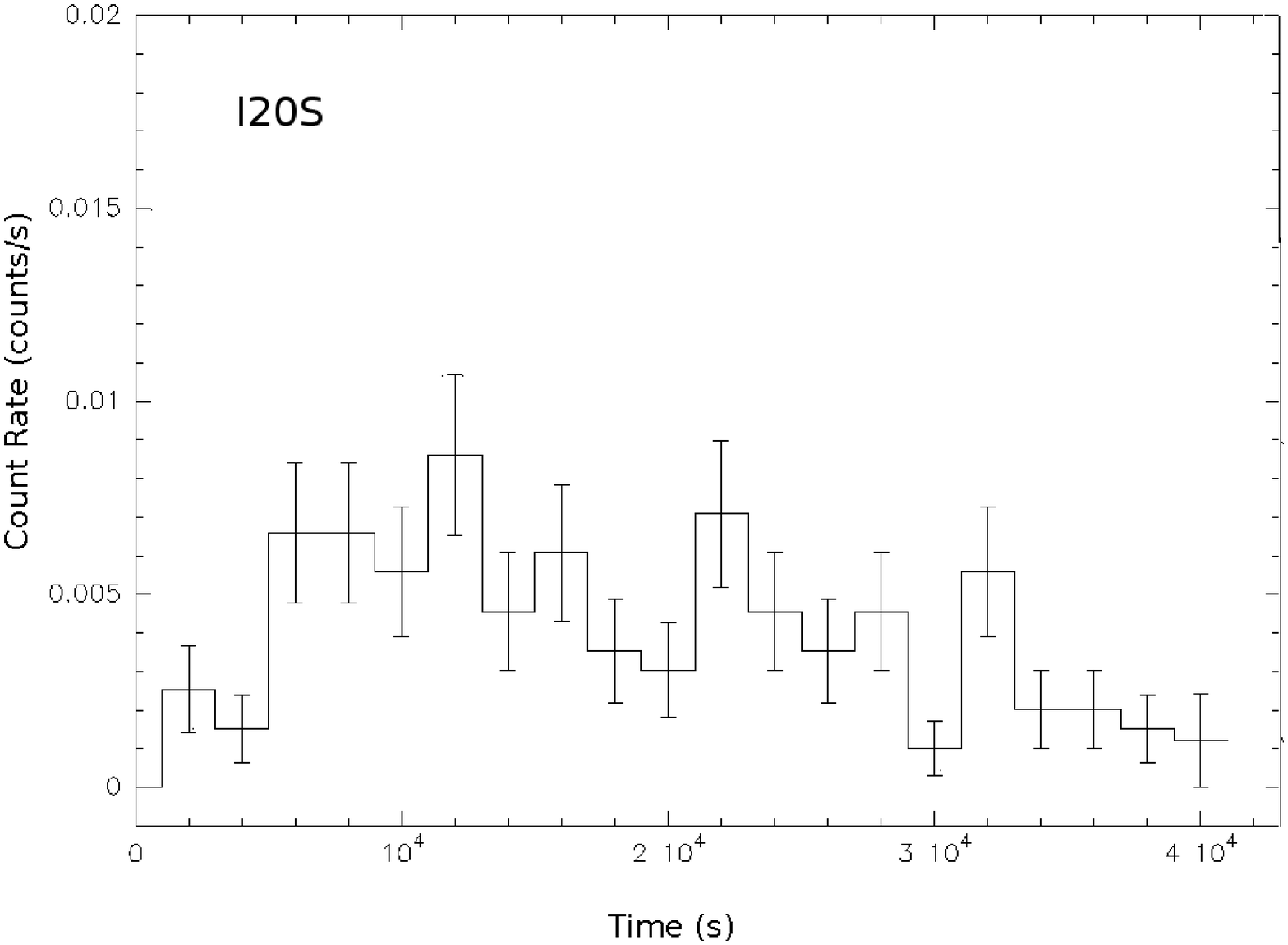}

\includegraphics{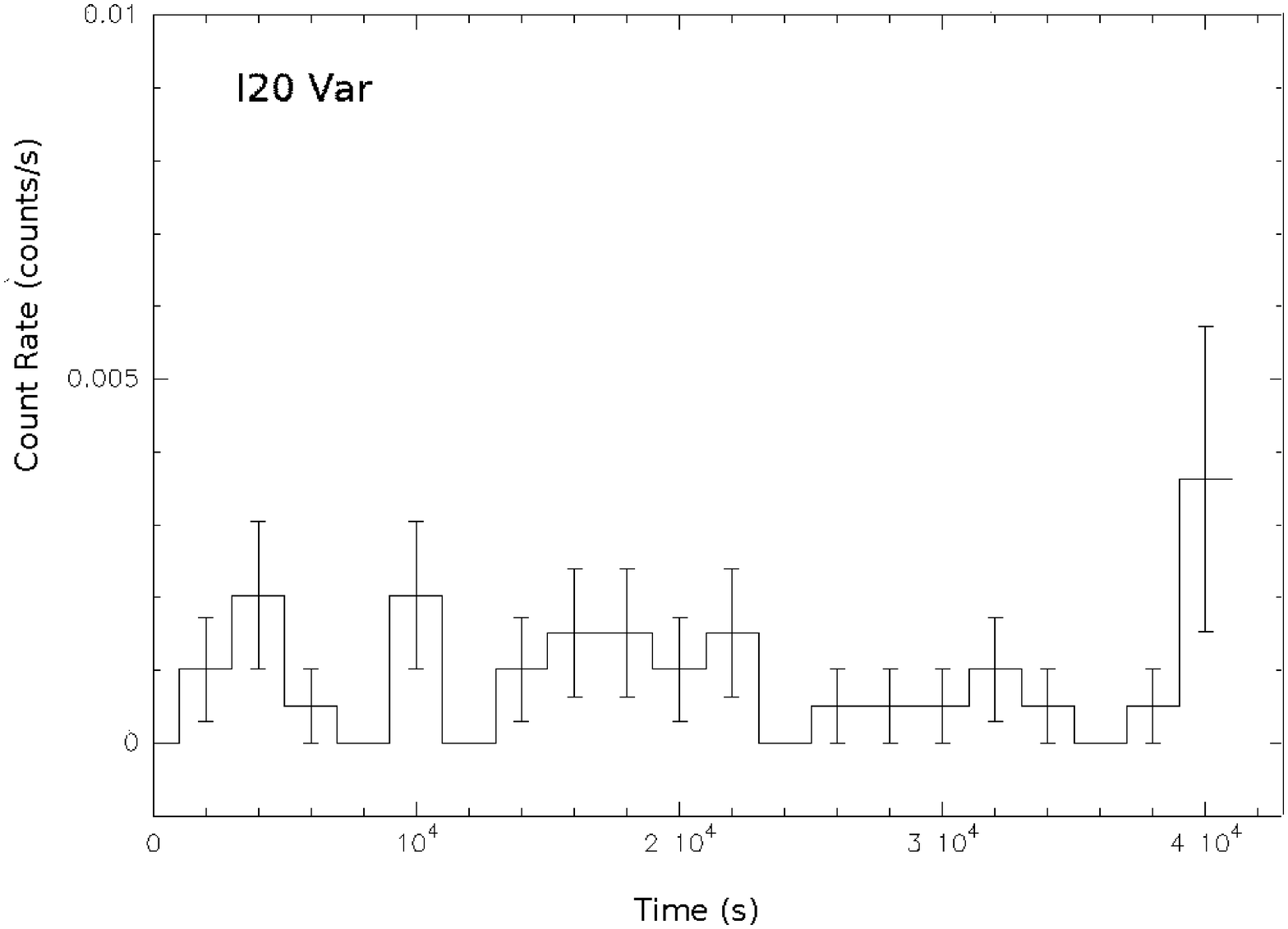}

\hspace*{0.0cm}
\begin{minipage}{15cm}
\vspace{16cm}
{\small Figure~3. {\it Top panel:} X-ray lightcurve of I20S. This source was found to be marginally variable ($\chi^{2}$ 
probability for constancy $1.1\times 10^{-5}$), but 
did not exhibit
any flare like behavior. {\it Bottom panel:} X-ray lightcurve
of the radio variable source I20Var. No X-ray variability was detected ($\chi^{2}$ probability for constancy 0.4).
The sudden rise at the end of the observations might indicate the onset of an
X-ray flare.}
\end{minipage}

\clearpage

\includegraphics{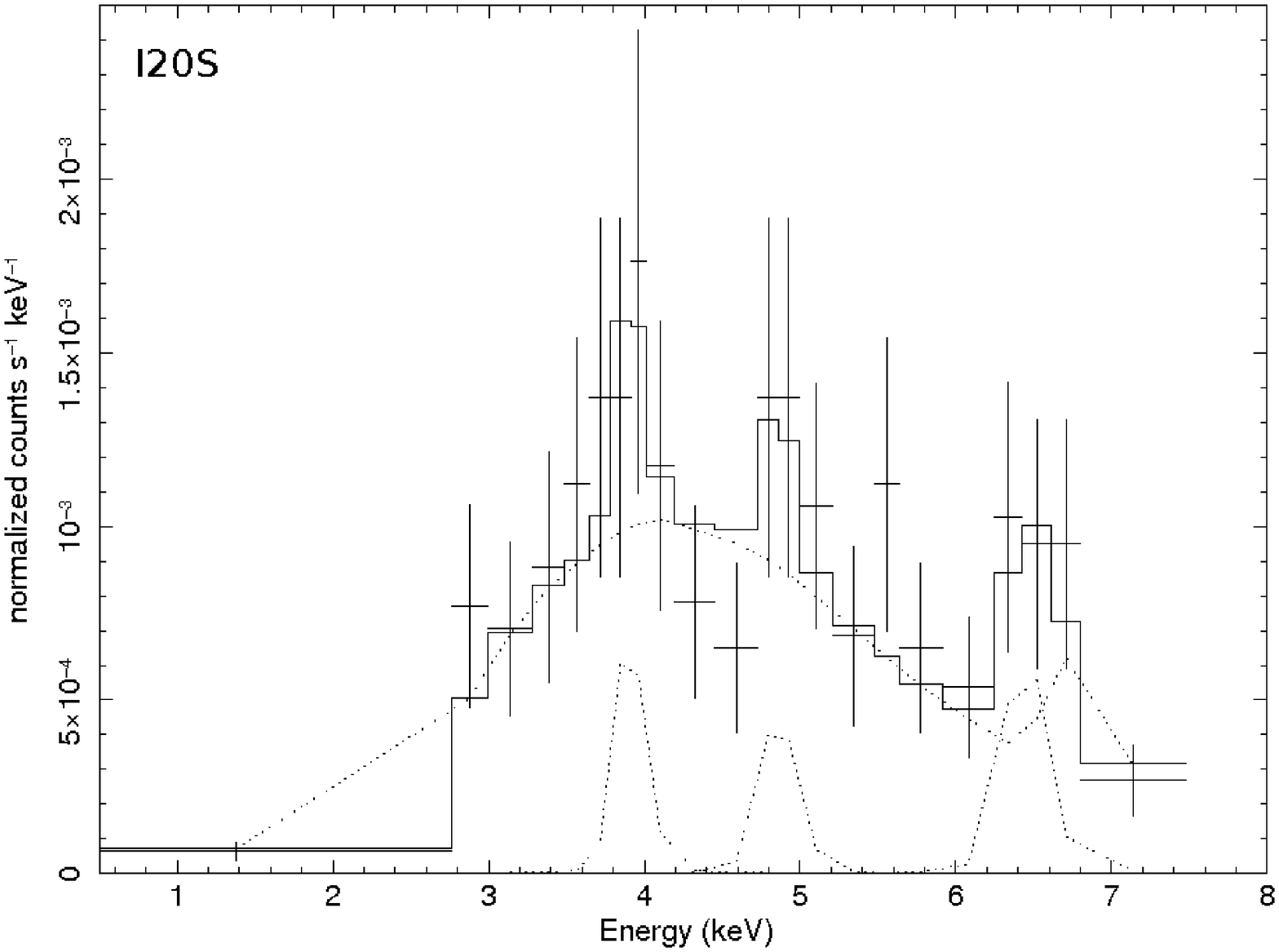}

\includegraphics{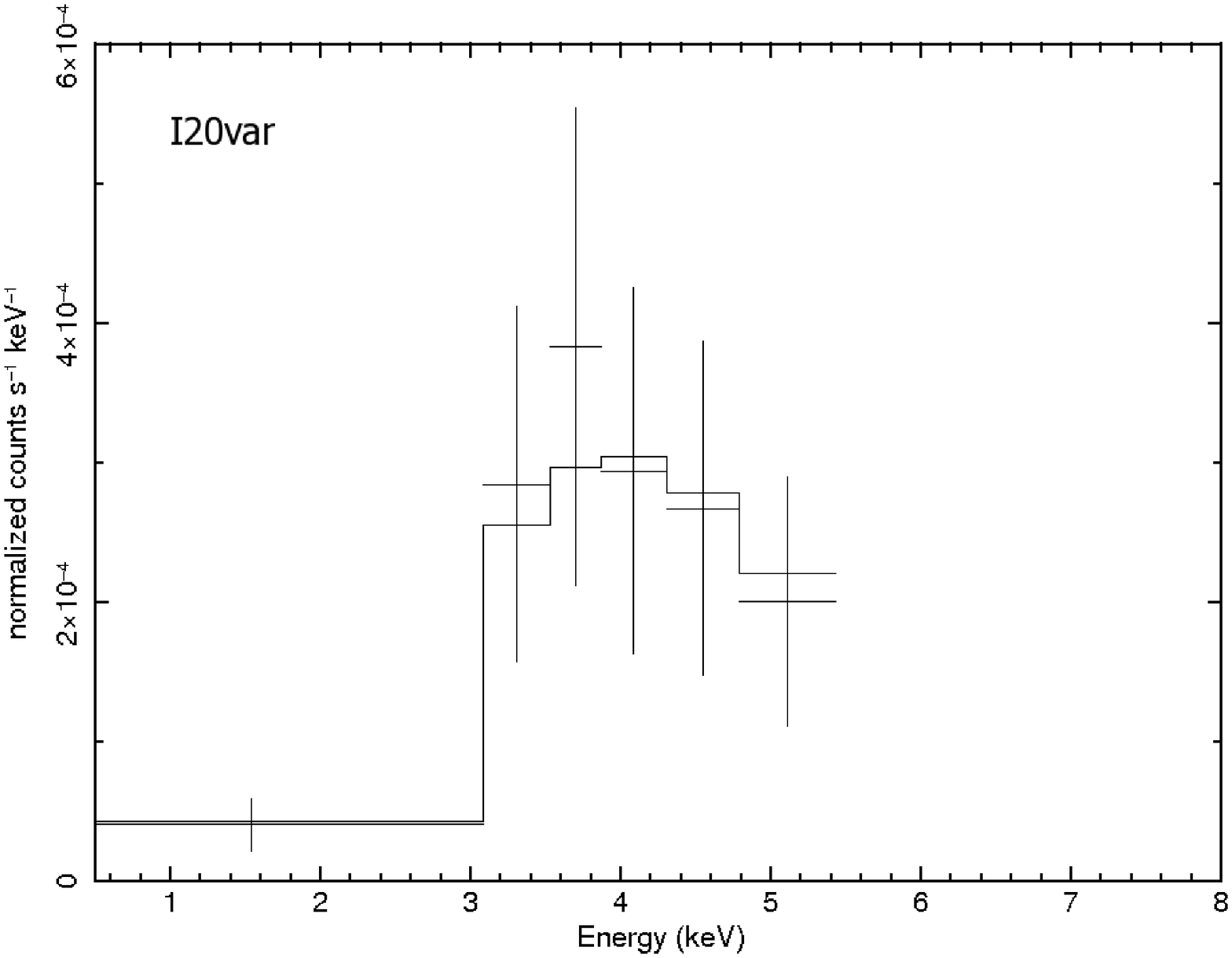}

\hspace*{0.0cm}
\begin{minipage}{15cm}
\vspace{16cm}
{\small Figure~4. {\it Top panel:} X-ray spectrum of I20S.
The  solid line shows a composite model consisting of a 1T APEC  model with an absorbing column of 
N$_H = 1.2^{+0.3}_{-0.2}\times 10^{23}\,$cm$^{-2}$, a plasma temperature of kT$=10.0^{+9}_{-2}\,$keV, 
and an abundance of 1.2 solar, plus 3 gaussians representing lines at energies of 3.9 and $4.9\,
$keV (Ca$\,$XIX) and $6.4\,$keV (Fe K$\alpha$). The dashed lines represent the individual models used.

{\it Bottom panel:} X-ray spectrum
of I20Var. The solid line shows a 0.2 solar abundance, 1T APEC model, with
an absorbing hydrogen column of N$_H = 1.1\times 10^{23}\,$cm$^{-2}$ and
a plasma energy of kT$=6.0\,$keV. 

}
\end{minipage}

\end{document}